\documentclass[twocolumn,twoside,slac_two,floatfix]{revtex4}
\usepackage{graphicx}
\usepackage{fancyhdr}
\begin{document}

\title{15~GHz Monitoring of Gamma-ray Blazars with the OVRO 40~Meter Telescope in Support of \emph{Fermi}}

%
\author{J.~L. Richards}
\author{W.~Max-Moerbeck}
\author{V.~Pavlidou}
\author{T.~J. Pearson}
\author{A.~C.~S. Readhead}
\author{M.~A. Stevenson}
\affiliation{California Institute of Technology, Owens Valley Radio Observatory, Pasadena, CA 91125, USA}

\author{S.~E. Healey}
\author{R.~W. Romani}
\author{M.~S. Shaw}
\affiliation{Department of Physics/KIPAC, Stanford University, Stanford, CA 94305, USA}

\author{L.~Fuhrmann}
\author{E.~Angelakis}
\author{J.~A. Zensus}
\affiliation{Max-Planck-Institut-f\"{u}r-Radioastronomie, Bonn 53121, Germany}

\author{K.~Grainge}
\affiliation{Astrophysics Group, Cavendish Laboratory, University of Cambridge, Cambridge, UK}

\author{G.~B. Taylor}
\affiliation{Department of Physics and Astronomy, University of New Mexico, Albuquerque, NM 87131, USA}

\begin{abstract}
  We present results from the first two years of our fast-cadence
  15~GHz gamma-ray blazar monitoring program, part of the F-GAMMA
  radio monitoring project. Our sample includes the 1158 blazars north
  of $-20^{\circ}$ declination from the Candidate Gamma-Ray Blazar
  Survey (CGRaBS), which encompasses a significant fraction of the
  extragalactic sources detected by the \emph{Fermi} Gamma-ray Space
  Telescope. We introduce a novel likelihood analysis for computing a
  time series variability amplitude statistic that separates intrinsic
  variability from measurement noise and produces a quantitative error
  estimate. We use this method to characterize our radio light
  curves. We also present results indicating a statistically
  significant correlation between simultaneous average 15~GHz radio
  flux density and gamma-ray photon flux.
\end{abstract}

\maketitle

\thispagestyle{fancy}

\section{INTRODUCTION}
The extragalactic gamma-ray sky is dominated by blazars, a class of
active galactic nuclei (AGN) that exhibit powerful, variable broadband
emission, likely due to observing a relativistic jet aimed close to
the line of sight.  Despite intensive study and modeling, the
mechanisms that generate the prodigious emission in these objects are
not well constrained.  Because of the broadband emission, simultaneous
multi-wavelength studies provide powerful probes of the structures and
processes that give rise to the blazar phenomena.

The Large Area Telescope (LAT) on the \emph{Fermi} Gamma-ray Space
Telescope (GST) provides unprecedented all-sky gamma-ray
monitoring, and detected 104 bright AGN with radio counterparts in its
first three months of operation and will likely detect many more as
its mission continues~\citep{abdo_bright_2009}.  This offers the
opportunity to perform simultaneous multi-wavelength studies of
statistically large numbers of sources.  Simultaneity of measurements
is important to eliminate ambiguities arising from variability.  Using
statistically large samples is also extremely important---although AGN
exhibit many common features, they make up a complex taxonomy and it
is difficult to extrapolate features from a small sample to AGN as a
whole.

Correlating gamma-ray emission with activity at radio wavelengths can
provide a powerful probe of blazar physics.  Through high-resolution
VLBI, radio observations are uniquely capable of spatially resolving
AGN and identifying the location of emission within the structure.  If
emission events at other wavelengths can be connected with events in
radio, their emission locations can perhaps be identified.  This is
extremely important to help constrain models of the emission and
eventually reach a full understanding of the AGN phenomena.

\subsection{OVRO 40~M Monitoring Program}
Since 2007, we have carried out 15~GHz observations of the 1158
sources north of $-20^{\circ}$ declination in the Candidate Gamma-Ray
Blazar Survey (CGRaBS) sample with the Owens Valley Radio Observatory
(OVRO) 40~M Telescope.\footnote{Data from the OVRO 40~M monitoring
  program are available from
  \url{http://www.astro.caltech.edu/ovroblazars}.}  The CGRaBS are a
sample of 1625 sources, mostly blazars, selected by their flat radio
spectra and X-ray fluxes to resemble the blazars detected by EGRET
\citep{healey_cgrabs:all-sky_2008}.  These were expected to constitute
a large sample of the extragalactic sources detected by \emph{Fermi}.
Each source is observed approximately twice per week with a thermal
noise floor of about 5~mJy. In addition to the CGRaBS core sample, AGN
detected by \emph{Fermi} are regularly added to our program if they
are found to be detectable with our sensitivity.

\subsection{The F-GAMMA Project}
The OVRO 40~M program is a key component of the \emph{Fermi}-GST
Multi-wavelength Monitoring Alliance (F-GAMMA)
project~\citep{fuhrmann_simultaneous_2007,angelakis_monitoringradio_2008}.
Also within the F-GAMMA project, spectra of a smaller sample of about
60 bright sources, selected \emph{ad hoc} based on historically
interesting behavior, are monitored monthly at twelve frequencies
between 2.7 and 270~GHz using the Effelsberg 100~m and Pico Veleta
30~m telescopes.  The coordinated F-GAMMA monitoring strategy provides
both broad spectral monitoring of a moderate number of sources, as
well as high-cadence monitoring of a very large, statistically
well-defined sample well-suited for population studies.

\section{QUANTIFYING VARIABILITY}
A number of measures of the time-independent variability amplitude of
a light curve are used in the AGN literature, such as the variability
index~\citep[e.g.][]{aller_pearson-readhead_1992} and the modulation
index~\citep[e.g.][]{fuhrmann_testinginverse-compton_2008}.  In
general, these methods produce a number that quantifies the amplitude
of variations in the source but do not produce an estimate of the
uncertainty in this number.  Furthermore, measurement errors in the
data, if considered at all, are not handled in a sophisticated way.
Finally, some of these methods suffer from pathological failures that
complicate interpretation. For example, in the presence of high noise,
the variability index can yield negative indices that must be
rejected.

\subsection{The Intrinsic Modulation Index}
To address these issues, we introduce the \emph{intrinsic modulation
  index} defined as $\hat{m}=\sigma_{0}/S_{0}$. Here $S_{0}$ and
$\sigma_{0}$ are the intrinsic mean and standard deviations of the
source flux density time series.  By ``intrinsic'' we mean excluding
contributions from errors in the measurement process.\footnote{This
  should not be confused with ``intrinsic behavior'' of an
  astronomical source---our method does not, for example, separate
  variability due to interstellar scintillation from variability in
  the source itself.}

Of course these intrinsic quantities are unknown, so we have
developed a likelihood method for estimating $\hat{m}$ from a measured
light curve and its associated measurement errors. For this method, we
postulate a parametric model for the intrinsic distribution of flux
densities from the source. Here we assume a Gaussian distribution
parameterized by its mean ($S_{0}$) and intrinsic modulation index
($\hat{m}$, a convenient rescaling of the standard deviation).  This
choice of distribution is not unique, and in fact is clearly not the
best description for many sources.  However, it is analytically
convenient and comparisons with other distributions suggest that the
errors that result are tolerable.

\begin{figure}
  \includegraphics[width=65mm]{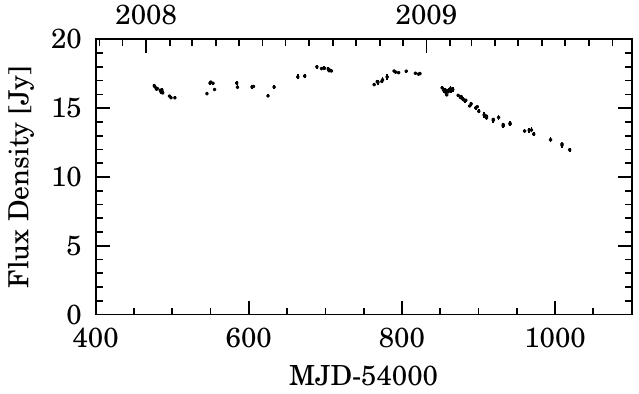}
  \caption{OVRO 15~GHz light curve for 3C~279.}
  \label{fig:lc_3C279}
\end{figure}

\begin{figure}
  \includegraphics[width=65mm]{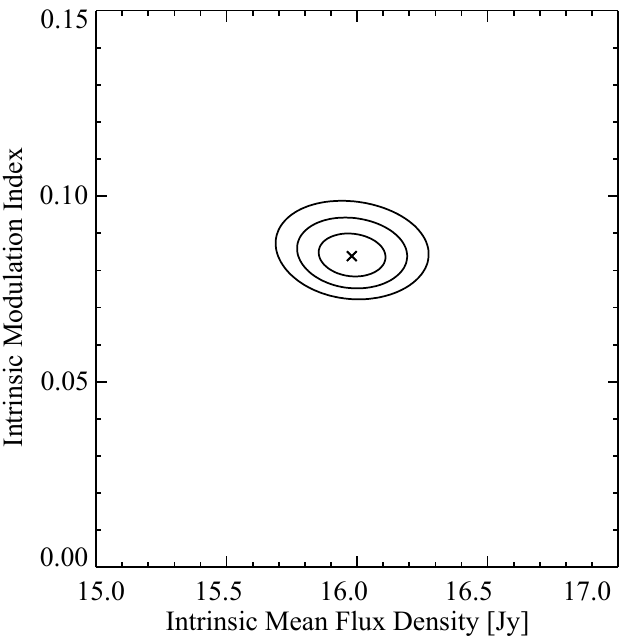}
  \caption{Maximum likelihood and 1, 2, and $3\sigma$ contours for 3C~279.}
  \label{fig:contours_3C279}
\end{figure}

\begin{figure}
  \includegraphics[width=65mm]{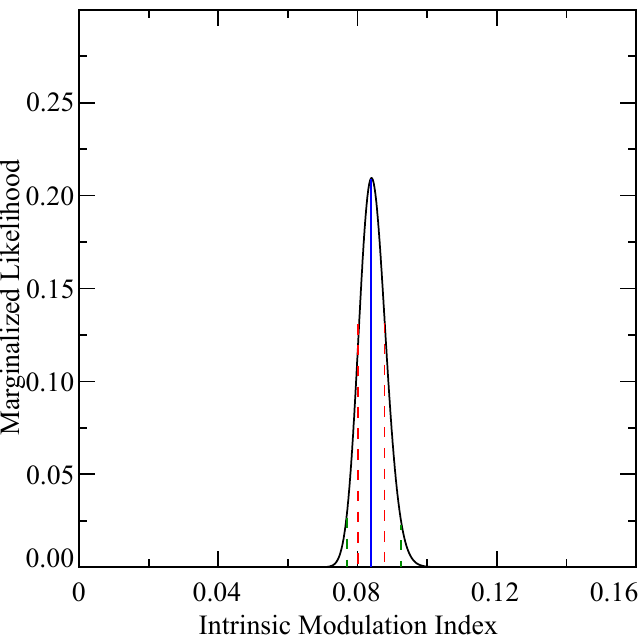}
  \caption{Marginalized likelihood distribution for the intrinsic
    modulation index for 3C~279.  Vertical lines indicate maximum
    likelihood value (sold) and $\pm 1$ and $\pm 2\sigma$ intervals
    (dashed).}
  \label{fig:imi_3C279}
\end{figure}

\begin{figure}
\includegraphics[width=65mm]{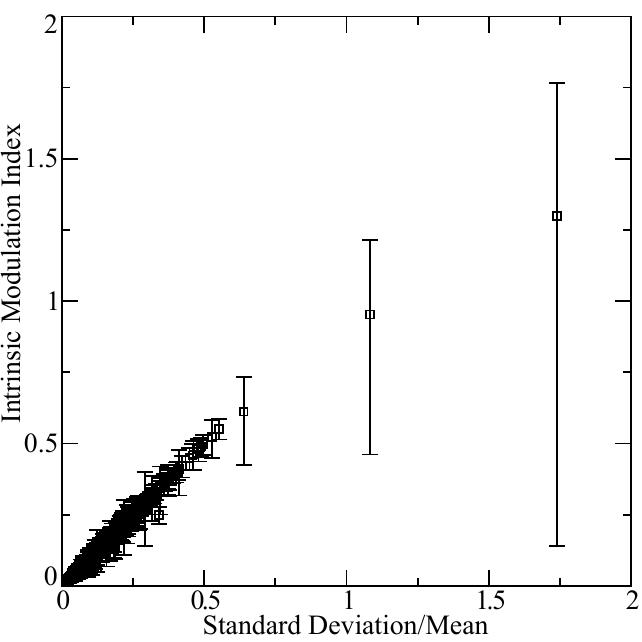}
\caption{Intrinsic modulation index for OVRO CGRaBS sample plotted
  against the traditional modulation index.}
\label{fig:sanity_check}
\end{figure}

Given a collection of flux densities and estimated measurement
uncertainties from a light curve (Figure~\ref{fig:lc_3C279}), we
compute the maximum joint likelihood of $S_0$ and $\hat{m}$ and find
the isolikelihood contours representing 1, 2, and $3\sigma$ (68.3\%,
95.5\%, and 99.7\%) confidences (Figure~\ref{fig:contours_3C279}).  We
then marginalize over $S_0$ to produce a likelihood distribution in
$\hat{m}$ (Figure~\ref{fig:imi_3C279}).  The peak of this distribution
gives our estimate for $\hat{m}$, and its width gives the uncertainty
(which is, in general, asymmetric).  The uncertainty includes the
contribution of measurement uncertainties, as well as effects of a
limited number of observations. This method agrees with the standard
modulation index, as shown in Figure~\ref{fig:sanity_check}, which is
reassuring.

\subsection{Population Studies}
\begin{figure}
  \includegraphics[width=65mm]{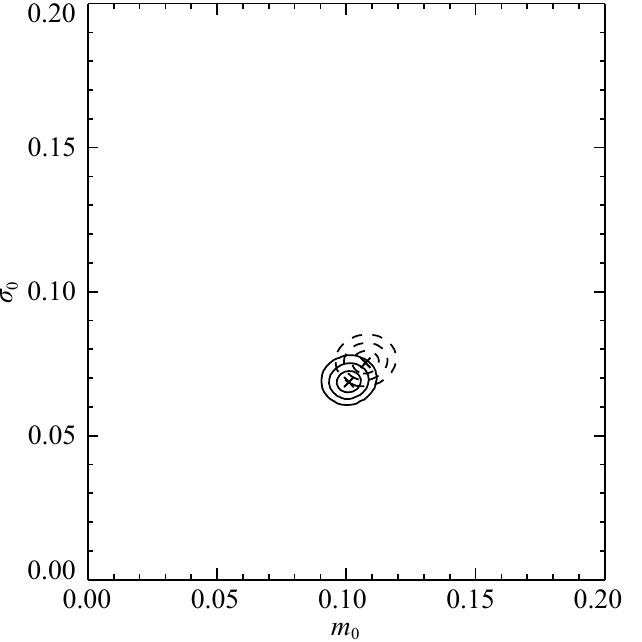}
  \caption{Likelihood contours for variability parameters of randomly split OVRO CGRaBS sample.}
  \label{fig:randomsplit}
\end{figure}

Armed with the intrinsic modulation index, we can compare the
variability amplitudes between subpopulations of our sample.  Again
using a likelihood analysis we model each subpopulation with a
distribution (here we show an example using a Gaussian distribution)
in intrinsic modulation index with mean $m_0$ and standard deviation
$\sigma_0$.  Figure~\ref{fig:randomsplit} shows the 1, 2, and
$3\sigma$ isolikelihood contours for a random split of our
population---as expected, these subpopulations overlap significantly
in parameter space.  This likelihood method permits two-dimensional
discrimination between population parameters, giving the capability to
find differences between populations with distributions that overlap
significantly in either parameter individually.

\section{RADIO-GAMMA CONNECTION}
A connection between radio and gamma-ray emission, manifested in a
correlation between flux densities in the two bands, would be
important evidence for common emission origins.  In the EGRET era,
some evidence for such a connection was found, but could not be
convincingly demonstrated to reflect an intrinsic correlation in
source luminosities~\citep{muecke_correlation_1997}.  A number of red
shift and sample selection effects can work to induce an apparent
correlation.  Recently, \citet{kovalev_relation_2009} found an
apparent correlation of \emph{Fermi} gamma-ray photon fluxes with
quasi-simultaneous 15~GHz VLBI core flux densities.  However, the
question as to whether this apparent correlation is significant in the
face of biases still stands.

\subsection{Significance: Likelihood Method}
We have developed a Monte Carlo method for evaluating the significance
of apparent correlations that accounts for both red shift and
selection effects and is compatible with sample selection methods that
cannot be quantified or reproduced.  Our method uses randomly-paired
luminosities and red shifts from the true data set to construct a
comparison sample with no intrinsic correlations. By constructing many
such samples, the probability density for chance correlations can be
estimated and used to evaluate the significance of the observed
correlation.

It should be noted that although this method can eliminate selection
effects when evaluating the significance of a correlation, it cannot
generalize a conclusion about the existence of a correlation outside
the sample being tested.  To make claims about, say, the general
blazar population, a representative subsample of that population is
still required.

\subsection{Results}
\begin{figure}
  \includegraphics[width=65mm]{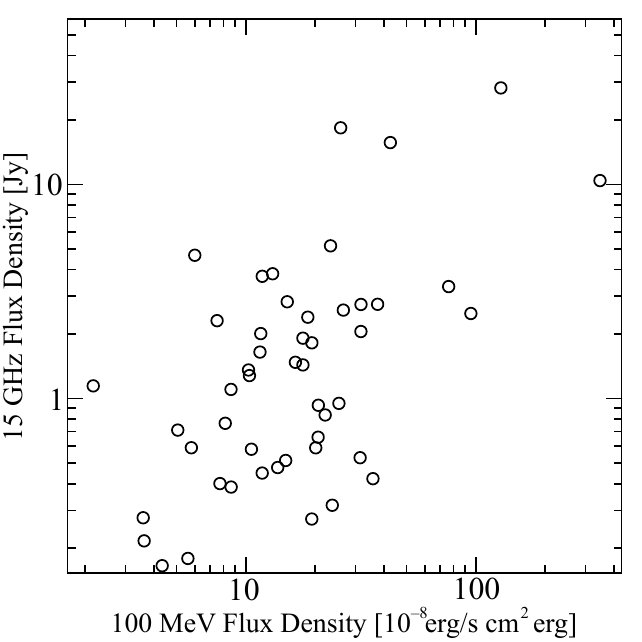}
  \caption{OVRO 15~GHz flux density versus \emph{Fermi}-LAT
    100~MeV flux density.}
  \label{fig:ovro_vs_fermi}
\end{figure}

There are 49 sources with known redshifts in both the OVRO CGRaBS
sample and the \emph{Fermi}-LAT 3-month bright AGN list
\citep{abdo_bright_2009}.  Simultaneous time average 15~GHz radio and
100~MeV \emph{Fermi} gamma-ray flux densities for each of these
sources are shown in Figure~\ref{fig:ovro_vs_fermi}.  These data are
averaged over the first three months of \emph{Fermi} operation,
August~4--October~30, 2008.  A correlation is visually apparent, and
the Pearson product-moment correlation coefficient for these data is
$r=0.56$.  Is this a significant correlation?

\begin{figure}
  \includegraphics[width=65mm]{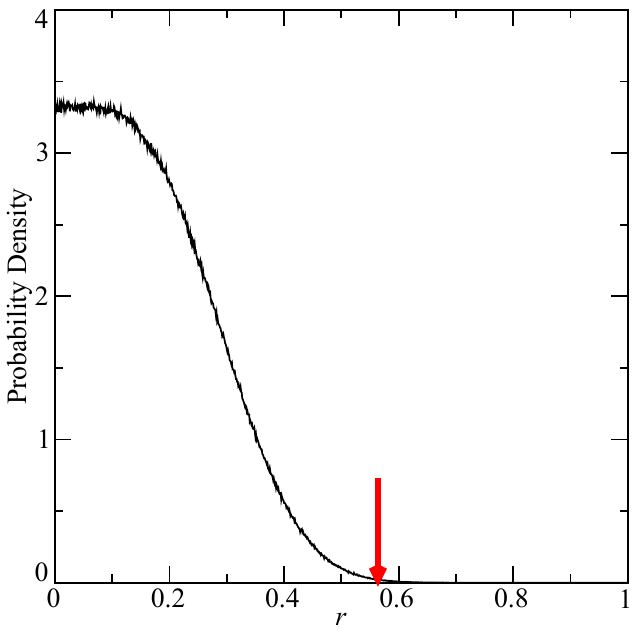}
  \caption{Monte Carlo-estimated probability density function for the
    correlation coefficient, $r$, between OVRO 15~GHz and \emph{Fermi}
    100~MeV flux densities.  Arrow indicates measured value.}
  \label{fig:ovro_vs_fermi_PDF}
\end{figure}

Applying our Monte Carlo method to the OVRO 15~GHz data, we obtain the
estimated probability density function shown in
Figure~\ref{fig:ovro_vs_fermi_PDF}.  We find a probability of about
$P=5\times 10^{-4}$ to find as large or larger a correlation by
chance.  Using multifrequency data from the F-GAMMA project for a
somewhat smaller population of sources we find a stronger correlation
($r=0.89$, $P=4\times 10^{-5}$) at 140~GHz and a decrease in both
correlation coefficient and significance as the radio frequency
decreases.  There is good agreement between the Effelsberg 14.6~GHz
result and the OVRO 15~GHz result, although the OVRO result is more
significant due to its larger sample size. Correlation results at all
frequencies are given in Table~\ref{table:correlations}.

\begin{table}[t]
\begin{center}
\caption{Radio-Gamma-ray Flux-Flux Correlation Results}
\begin{tabular}{cccc} \hline
  \textbf{Frequency} & \textbf{Correlation} & \textbf{Chance} & \textbf{Site\footnote{Site legend: PV=Pico Veleta IRAM 30~m; EFF=Effelsberg 100~m; OVRO=OVRO 40~m.}}\\
  \textbf{[GHz]} & \textbf{Coefficient} & \textbf{Probability} & \\ \hline
  142 & 0.89 & $4\times 10^{-5}$ & PV \\ 
  86  & 0.86 & $2\times 10^{-5}$ & PV \\ 
  43  & 0.83 & $7\times 10^{-4}$ & EFF \\ 
  32  & 0.74 & $6\times 10^{-4}$ & EFF \\ 
  22  & 0.59 & 0.01 & EFF \\ 
  15  & 0.56 & $5\times 10^{-4}$ & OVRO \\ 
  14.6 & 0.49 & 0.03 & EFF \\ 
  10.5 & 0.43 & 0.05 & EFF \\ 
  8.4 & 0.40 & 0.06 & EFF \\ 
  4.8 & 0.40 & 0.08 & EFF \\ 
  2.6 & 0.43 & 0.06 & EFF \\ \hline
\end{tabular}
\label{table:correlations}
\end{center}
\end{table}

\section{CONCLUSIONS}
Densely-sampled light curves for the 1158 northern CGRaBS and a number
of other sources have been collected using the OVRO 40~M Telescope
since 2007.  In addition, monthly radio spectra of about 60 sources
have been collected through the F-GAMMA project.  Work is underway to
combine these radio light curves with gamma-ray data from the
\emph{Fermi}-GST.

We have developed a new likelihood method, the intrinsic modulation
index, for quantifying the degree of variability in a light curve.
Unlike standard measures, our method quantifies the uncertainty in its
output after accounting for measurement errors, and can be used to
compare data sets that are inhomogeneous with respect to measurement
errors or number of measurements.

Finally, we conclude that we have found clear evidence for a
statistically significant correlation between OVRO and F-GAMMA radio
and \emph{Fermi}-LAT gamma-ray flux densities.  Using our Monte Carlo
method, we have demonstrated that this correlation is not the result
of red shift or selection effects.  Our method can be applied to
samples, such as the F-GAMMA sample, that were not selected using
statistical criteria.  This correlation is strongest and most
significant at higher radio frequencies, but is still highly
significant at least as low as 15~GHz.

\bigskip 
\begin{acknowledgments}
  The OVRO 40~M program is supported in part by NASA grant NNX08AW31G
  and NSF grant AST-0808050.  WM acknowledges support from the
  U.S. Department of State and the Comisi\'on Nacional de
  Investigaci\'on Cient\'ifica y Tecnologica (CONICYT) in Chile for a
  Fulbright-CONICYT scholarship.  VP acknowledges support for this
  work provided by NASA through Einstein Postdoctoral Fellowship grant
  number PF8-90060 awarded by the Chandra X-ray Center, which is
  operated by the Smithsonian Astrophysical Observatory for NASA under
  contract NAS8-03060.
\end{acknowledgments}

\bigskip 
\bibliography{jlrbib}

\end{document}